# Large Area Growth of Aligned CNT Arrays on Spheres: Towards the Large Scale and Continuous Production[**]

By Rong Xiang, Guohua Luo, Weizhong Qian, Yao Wang, Fei WEI[*], Qi Li

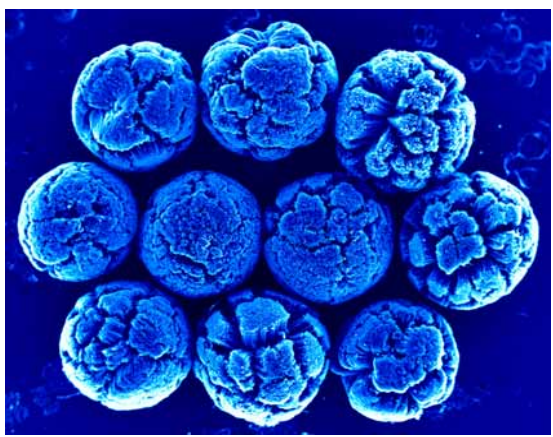

Table of Contents: A novel strategy for the large scale and continuous production of aligned carbon nanotube arrays using millimeter-diameter spheres as growth substrates is reported. The present technique is more productive than the conventional process on flat wafers because of the higher available growth surface and the good fluidity of the spherical substrates. It can be adapted for the industrial production and application of aligned carbon nanotube arrays with lengths up to millimeter.

[*] Mr. R. Xiang, Dr. G. H. Luo, Dr. W. Q. Qian, Dr. Y. Wang, Dr. F. Wei, Beijing Key Laboratory of Green Chemical Reaction Engineering and Technology, Department of Chemical Engineering, Tsinghua University, China. Dr. Q. Li, Cnano Group Limited, Hong Kong, China. To whom correspondence may be addressed. Weifei, weifei@flotu.org, Department of Chemical Engineering, Tsinghua University, Beijing 100084, China; tel, 86-10-62788984; fax, 86-10-62772051. Current address for R. Xiang: Department of Mechanical Engineering, The University of Tokyo, 7-3-1 Hongo, Bunkyo-ku, Tokyo 113-8656, JAPAN.

[**] The work was supported by China National '863' Program (No. 2003AA302630), China National Program(No. 2006CB932702), NSFC Key Program (No. 20236020), FANEDD (No. 200548), Key Project of Chinese Ministry of Education (No. 106011), THSJZ, and National center for nanosicence and technology of China (Nanoctr). Supporting Information is available online from Wiley Inter-Science or from the author.

Since the first synthesis in 1996,[1] vertically aligned carbon nanotube (CNT) arrays have become one focus in nanoscience and nanoengineering. These aligned CNTs in the forest-like arrays have many attractive properties, such as identical tube length, uniform orientation, extra high purity, easy spinning into macroscopic fibers, etc. Therefore, the as-grown carpet can be used directly to construct field emission devices, anisotropic conductive materials, and multi-functional membranes.[2-6] Even after the loss of the original alignment, longer and straighter multi-walled CNTs from the arrays were found to be better than randomly aggregated multi-walled or even single-walled CNT in improving the electronic, mechanic and thermal properties of the polymers.[7] However, unlike the randomly oriented CNTs aggregates that have already been scaled up by Hyperion Catalysis International, Rice University, etc., the synthesis of CNT arrays is still only performed on the lab scale of less than one gram on inch-sized silicon, quartz or glass wafers.[2, 8, 9] This has low production efficiency, and a lot of other promising applications that need larger amounts of CNTs can hardly be carried out and aligned CNTs can never been finally commercialized in bulk quantity.

To produce more aligned arrays, one approach is to improve CNTs Length.[10, 11] This is relatively difficult because, in most of the cases, the CNT arrays stop growing within several millimeters while the deactivation mechanism is not yet clearly understood. Since the CNT array yield is proportional to the size of the growth region (i. e. the size of the substrate with effective catalyst), another way, which is more controllable and promising, is to use a larger growth area as described by Singh et al.[12] As another consideration, a continuous operation is preferred to a batch operation because of higher efficiency and lower cost. In a word, large-area growth and continuous transport cannot be easily achieved on conventional flat wafers.

Spheres were seldom considered as a promising growth substrate candidate, because usually well aligned CNT arrays cannot form without a flat surface, e.g. on sub-micron or several micron silica particles with large curvatures as demonstrated previously.[13,14] However, once the diameter of a sphere exceeds several hundreds micrometer, the curvature of the spherical surface is negligible at the small area where CNTs can grow and align (usually hundred μm$^2$ for multi-walled CNT and several μm$^2$ for single-walled CNT as demonstrated in the patterned growth[15,16]). In other words, there is no reason why CNTs can not grow vertically and form well aligned array on spheres with larger diameters. The most attractive advantages of spheres are: first, spheres can be easily stacked yet can provide much larger outer surfaces than conventional flat wafers; second, the spheres have the best fluidity among all shapes of solid particles. Here, we present the successful bulk synthesis of aligned CNT arrays on large amounts of spheres. This approach is scalable, continuable and cost-effective. Specifically, when cheap inorganic oxide spheres with diameters around 1 mm and total volume of 1 dm$^3$ were introduced as the growth substrates in ferrocene pyrolysis, the surface for the CNT growth was as large as 5 m$^2$ (equals to that of 10000 pieces of one-inch wafer). Vertically aligned CNT arrays grew perpendicularly on all these spheres at the same time, and 100 g product was obtained in 30 min. Furthermore, these spheres and CNTs with an average diameter of 1.5 mm still have excellent fluidity and can be moved easily. The whole process can be operated continuously at low cost.

When grown on a flat quartz substrate, the CNTs were usually straight and perpendicular to the substrate because the growth was restricted by surrounding CNTs into the same direction. On the spherical substrates, the products obtained from a 30min-reaction were still spherical particles but with larger diameters as shown in Fig. 1a. After we peeled off some of the carbon, CNTs were found to have grown vertically all over the sphere as shown in Fig. 1b. High magnification SEM and TEM micrographs, Fig. 1c and 1d, confirmed the high degree of alignment. The thickness of the CNT carpet was 300 μm and the average diameter was about 35 nm, which are both similar to the arrays grown on a flat quartz. Besides, these CNTs also have almost the same purity, characterized by TGA, and the same defect degree, characterized by Raman spectroscopy, as the plane-grown CNTs (see supporting information). These similar characteristics can be attributed to the consistence of a sphere and a flat wafer in small regions as mentioned above.

However, as the entire array on a single sphere is curved into a spherical membrane, the CNTs at the bottom part of a sphere are always growing under the pressure of the heavy core and have to lift it gradually to keep the growth. This is a unique phenomenon for spheres and never exists on the conventional flat wafers. Viewing a particle from three mutually perpendicular directions, all the CNTs on the entire sphere are nearly the same in length (see supporting information). Such symmetrical structure confirmed that the growth driving force in local region (bottom of sphere) is large enough to lift an object that is thousands times heavier than the CNTs themselves. A rough estimation shows that the pressure on the bottom CNT array caused by single sphere is about 20Pa. It is one order of magnitude smaller than the previously reported pressure extent (250Pa),[17] against which multi-walled CNT arrays cannot keep growing without obvious influence of pressure. Taking into account the larger average tube diameter in our case, the pressure limit that our aligned arrays can sustain is probably even higher. Therefore, the CNTs can raise at least 12 of such spheres when growing vertically. This strong driving force for the vertical growth is critical in the following mass production using large amounts of these inexpensive spheres stacked together.

After the successful growth of well-aligned and high-quality CNTs on spheres, we scaled up this process in a larger reactor using 1 dm$^3$ of such 1mm-spheres stacked in about 1cm high. As even the densely stacked spheres are of high porosity rate, the feedstock can easily diffuse and reach the bottom area through the continuous channel among the spheres. In the mean time, the above mentioned strong driving force during the CNT growth ensured that even the underlying spheres can grow into arrays against the presence of above spheres. Thus, all the spheres can be utilized as growth substrates. The average diameters of the spherical particles increased from about 800 μm (before the reaction) to about 1300 μm (after the reaction) and the total pack bed volume expanded several times as shown Fig. 2 (also see supporting information). Because the surface area (about 5 m$^2$) is four orders of magnitude higher than the conventional 1 inch flat substrate, the weight of all aligned multi-walled CNTs synthesized on these spheres in 30 min is nearly 100 g. To our knowledge, this is the largest output of well aligned CNT arrays. Meanwhile, these CNTs are also similar to the conventional ones on flat wafers in quality.

Furthermore, it was observed that the as-grown spheres covered with CNT arrays with lengths up to several hundreds micrometer still had very good fluidity. They can move freely without injury to the aligned structure, as was seen by tilting the glass dish (see supporting information). The response angles (i.e. the angle at which the spheres start to flow down an incline) of the spheres before and after the CNT growth were measured to be 17 and 27 (+-2), respectively, indicating the good fluidity of the spheres even after CNT growth. Such excellent fluidity cannot be obtained with any shaped particles other than spheres. This is important because it allows the transfer of the products out of the hot reactor by some simple methods (e.g. via a jet of inert gas). After addition of new spheres in the reactor in situ, the CNT synthesis can be kept continuously without stopping and cooling down the reaction, which can further promote the production and lower the cost significantly (see supporting information). Thus, one can be optimistic that a kilogram scale can be reached at a low cost in the very near future as using the design and procedure in our patent.[18] It is worthwhile noting that the stable structure does not mean a difficult removal of the CNTs from the spheres. The separation of the products and the substrates can be easily achieved by sonication (for the applications where the original alignment is unnecessary) or by gas milling (for the applications where the original alignment should be maitained), as shown in Fig. 3. In the case of gas milling, as the CNT-CNT interaction in the array is much stronger than the CNT-substrate interaction, the force generated during gas milling was able to separate the CNTs from substrates while keeping the CNTs almost in their original alignment. This technique is suitable for large scale separation in the future and the spheres after the removal of carbon can even be used for a second CVD in a recyclable way.

Interestingly, the morphology of the aligned CNT arrays depended on the CNT length. When the thickness of CNT carpet reached about 1/3 of the sphere diameter, the continuous carpet began to develop cracks due to the increasing outer surface area and consequently the increasing tension force on the array top. Without the constraint imposed by adjacent CNTs, the degree of alignment decreased gradually, as demonstrated in Fig. 4. Accordingly, there is a relationship between the curvature (sphere size) and the CNT length that can be grown without cracking. Therefore, in the case of too small spheres as demonstrated in previous work, the surface of submicron sphere is too much curved to grow long aligned CNT arrays.[13,14] In the other case, spheres that are too large will also have problems. Firstly, the available outer surface area per volume is less, and therefore the products yield less. Secondly, too high pressure on the CNTs caused by the gravity force of spherical substrates may hinder the successful growth (the CNT arrays need to lift the larger and heavier spheres above them while growing). Our current chosen sphere size (around 1 mm in diameter) is suitable for the mass production of sub-mm CNT arrays.

It is also worthy noting that on our ceramic spheres comprising $SiO_2$, $Al_2O_3$ and $ZrO_2$, the support site for the *in situ* formed iron catalyst is $SiO_2$, since it was found that well aligned CNTs can neither be formed efficiently on pure $Al_2O_3$ nor $ZrO_2$ under the same conditions (see supporting information). The reason why we chose this composite material is that these spheres are cheap and commercially available, which is critical for an industrial application.

The present successful large-area synthesis of vertically aligned array on spheres will make it possible to carry out research on applications that need a large amount of CNT arrays, such as fabricating multi-functional composite material. We have confirmed that the long CNTs arrays are highly efficient for constructing conducting matrixes in a polymer and also on a flat substrate. For a CNT-polymer composite with a specific electric conductivity, the needed loading of such long and straight CNTs is always much lower than our previous shorter and more curved ones from the aggregations.[19] The CNT arrays were also found more effective for applications in electromagnetism shielding. The detailed results will be presented in our later reports.

In summary, a large scale production technique for aligned CNT arrays was presented. By using spheres as novel growth substrates in a conventional and laboratory scale floating catalyst reactor, 100 g of aligned multi-walled CNT arrays was produced in 30 min. The process can be made continuous due to the solid structure and good fluidity of the spheres. The present work can provide the foundation for large scale applications of these long and well aligned CNT arrays.

*Experimental*

CNT arrays were fabricated via floating CVD, similar to our previous work.[20] Typical experiments were performed in two reactors, one horizontal quartz tube (No. 1, 25 mm in diameter and 1200 mm in length) and the other larger horizontal stainless steel reactor with quartz inner tubes (No. 2, 200 mm in diameter and 1500 mm in length). Ferrocene was used as the catalyst precursor and cyclohexane was used as the carbon source. The concentration of ferrocene in cyclohexane was 20 g/L. The substrates were commercially available ball-milling spheres consisting of 50% $SiO_2$, 30% $Al_2O_3$ and 20% $ZrO_2$. The spheres, sometimes along with a quartz substrate as reference, were first put in the central part of the tube and heated to 800 °C. The feedstock solution was then injected by a motorized syringe pump into the reactor with an atmosphere of 90% Ar and 10% $H_2$. The feed rate of the feedstock and flow rate of the carrier gas were 5 ml/h and 600 ml/min for reactor No. 1, 1000 ml/h and 120 l/min for reactor No. 2. The 100 g product was obtained in the No. 2 reactor in 30 min. All the products were characterized by the scanning electronic microscopy (JSM 7401, excited at 3 kV), the transmission electron microscopy (JEOL2010, excited at 200 kV), the Raman Spectroscopy (RM2000, Renishaw, excited at 514.5 nm) and thermal gravimetric analysis (TGA, TA2050, heated at 20°C/min).

Figures:

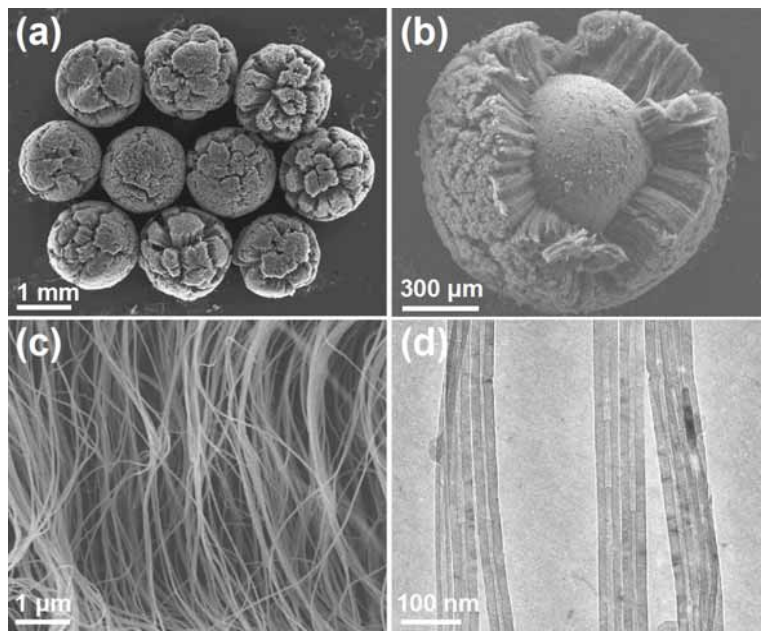

**Fig. 1.** (a), (b), (c) SEM and (d) TEM micrographs of the as-grown multi-walled CNT arrays vertically aligned on the spherical surface. The as-grown CNT-sphere is of great similarity to the phoenix-tree ball which enlighten us for the present work (see supporting information).

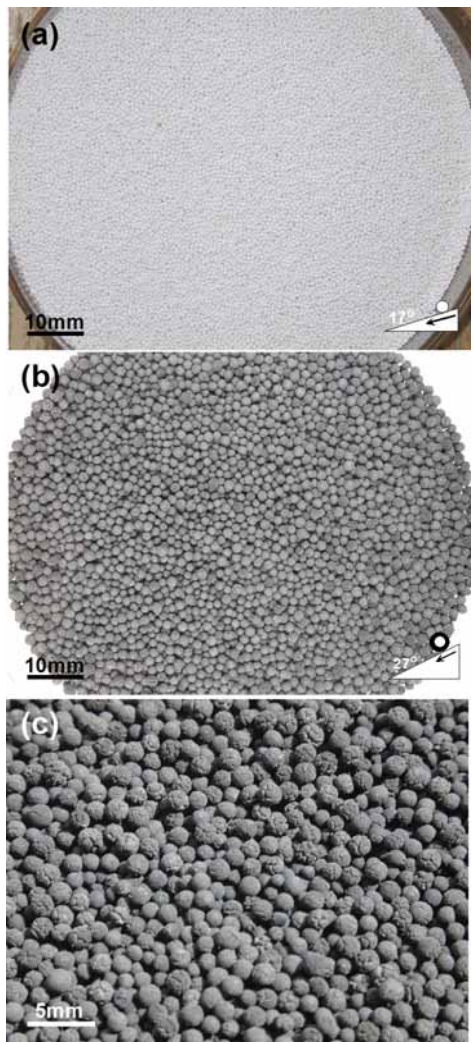

**Fig. 2.** Bulk production of well aligned MWNT arrays on large amounts of spheres: (a) spheres before the reaction, (b) spheres after of reaction of 30 mins, (c) enlarged image of as-grown spheres. The diameter of the sphere increases and the volume expands after the reaction.

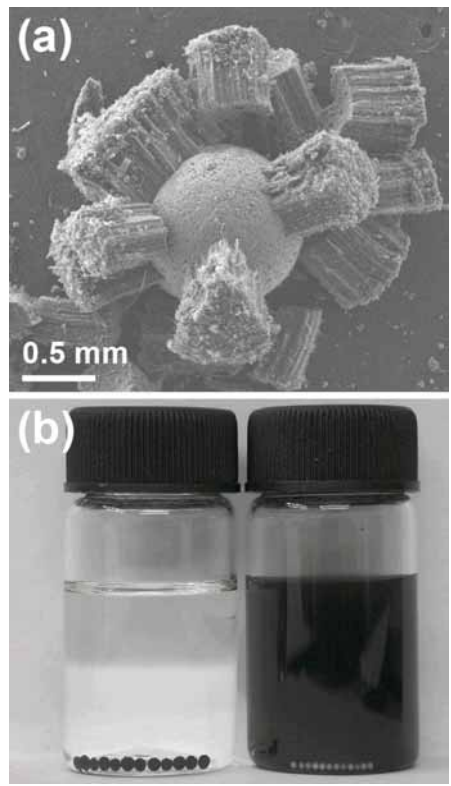

**Fig. 3.** Two easy methods to separate as-grown CNTs from the particles. (a) SEM micrograph of the aligned CNT arrays and spherical substrate structure destroyed by jet milling; (b) the spheres in ethanol before and after keeping them in a sonic cleaner for 10 seconds.

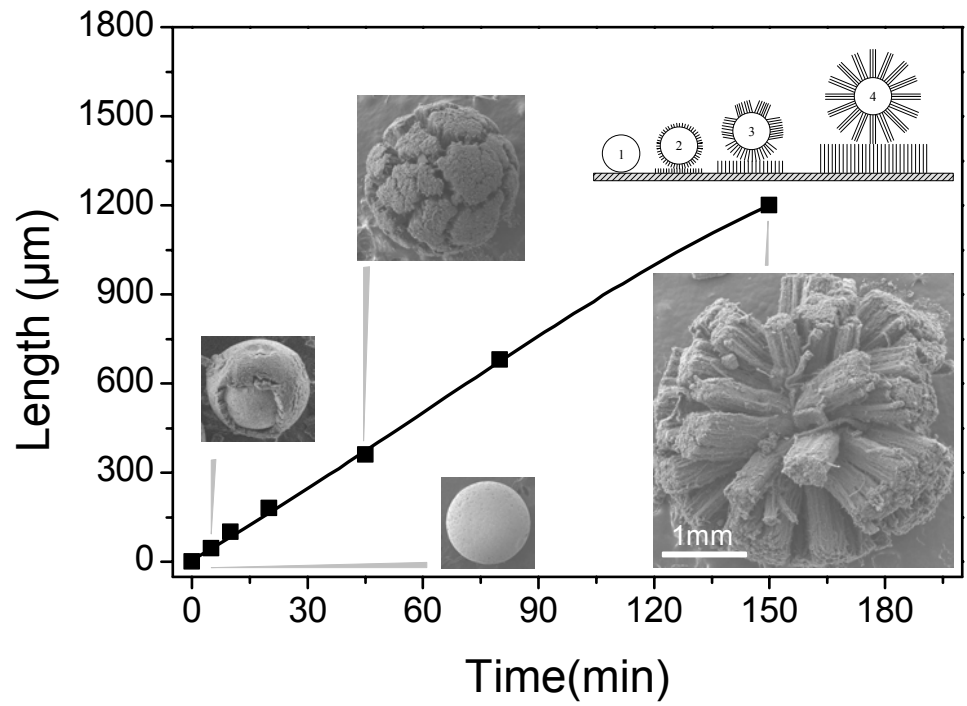

**Fig. 4.** Effect of growth time on the lengths and morphologies of the as-obtained CNTs on spheres. Insets are SEM micrographs of from different time growth and a schematic diagram of the growth process(top-right).

*Supporting information*

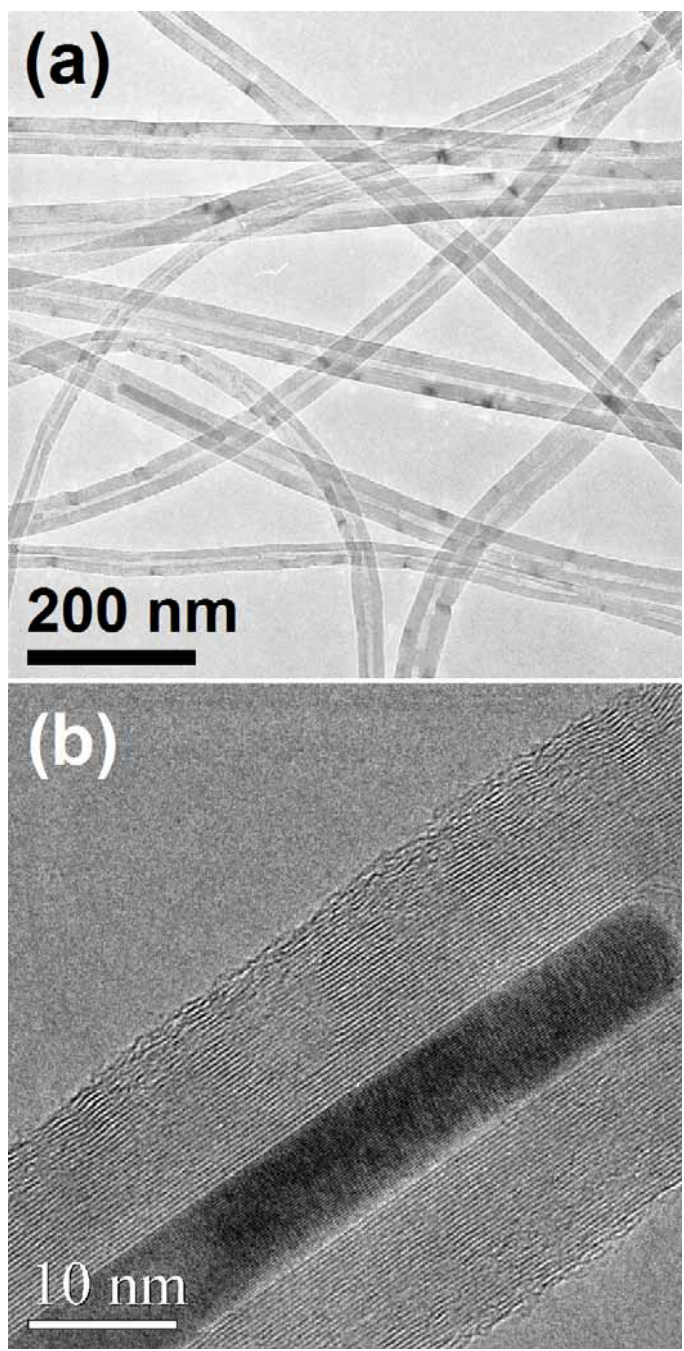

**Fig. S1.** (a) TEM and (b) HRTEM micrographs of CNTs grown on spheres. The MWNTs are well graphitized and free of other carbon materials. The only impurity is the 3% catalyst particles (usually as gamma-Fe or $Fe_3C$) encapsulated inside the channels of the CNTs.

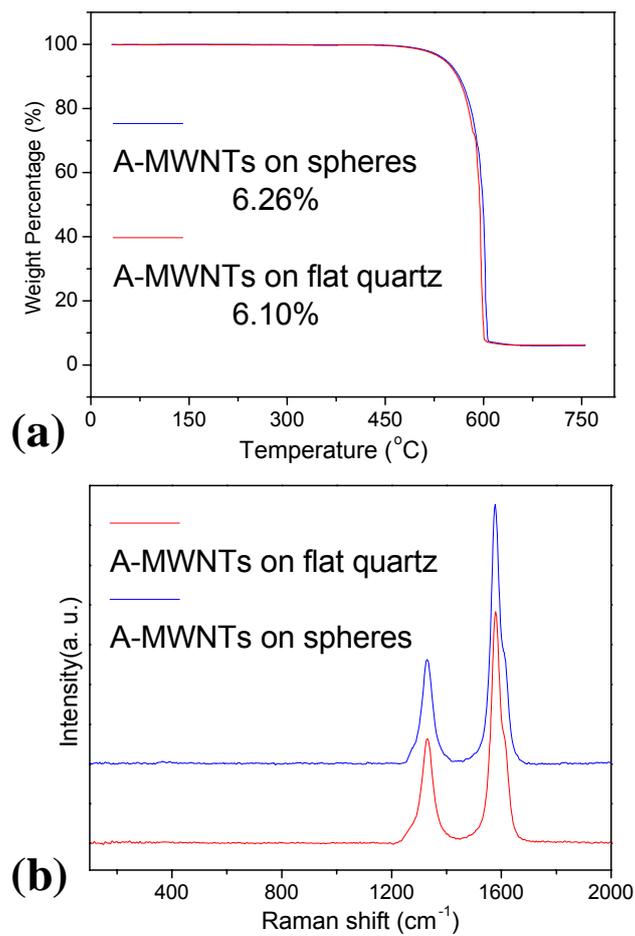

**Fig. S2.** (a) TGA and (b) Raman spectroscopy of CNT arrays. The products from the spheres are of the similar purity and defect degree to those from the conventional flat quartz wafers.

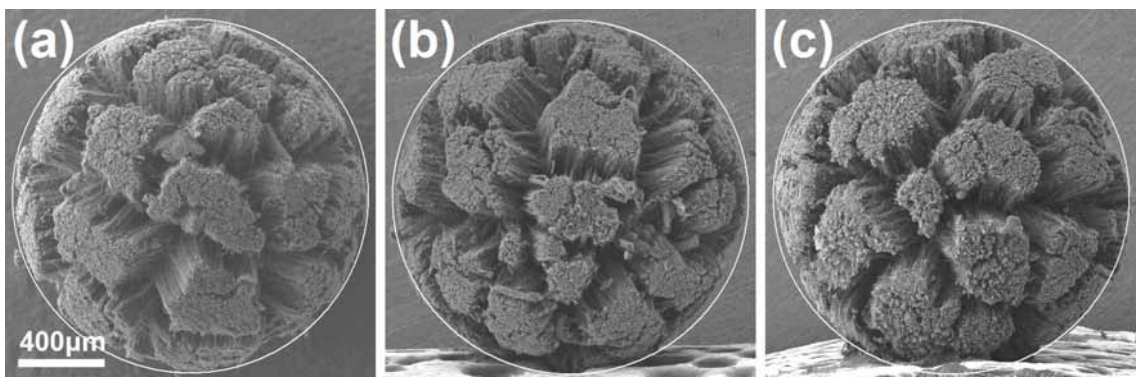

**Fig. S3.** SEM images from different directions of a sphere with as-growth CNT arrays: (a) top view, (b) front view and (c) side view. Three white circles indicate that the sphere with CNT arrays is still in a perfectly spherical shape and thus verify that aligned CNTs grown along all the directions of a sphere are almost the same in length.

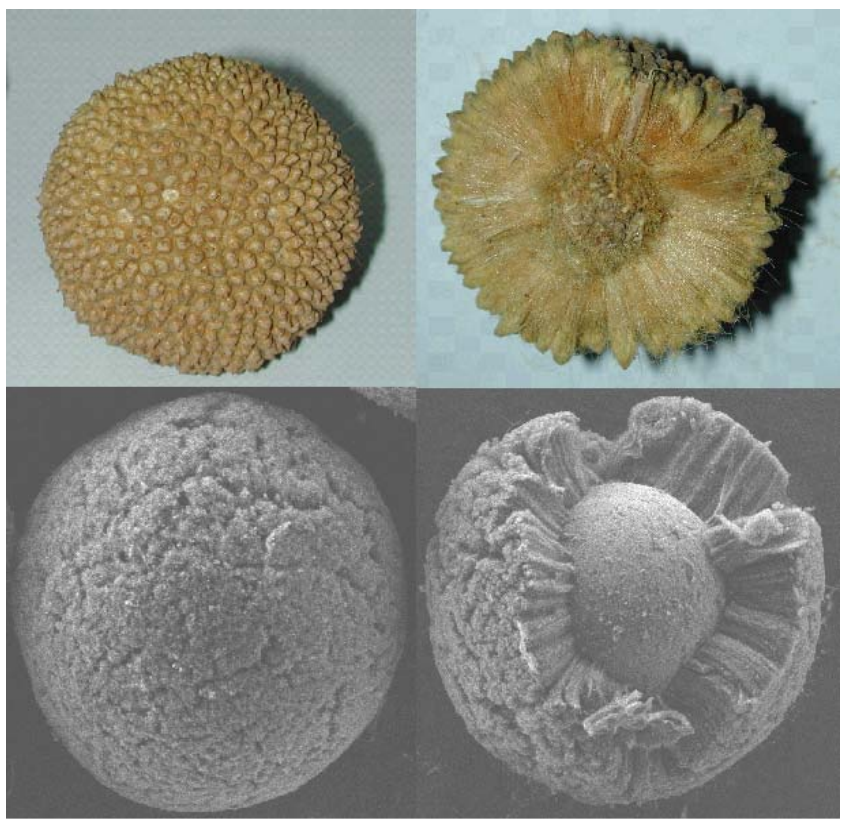

**Fig. S4.** Nature-existing balls from the phoenix tree and as-grown CNT-substrate spheres from our experiment. These two structures are of great similarity and the present work is mainly enlightened by this structure.

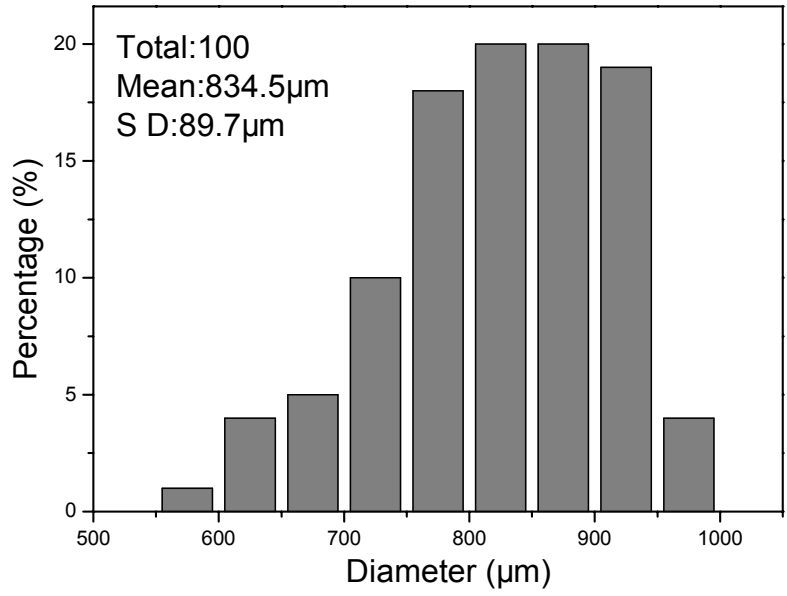

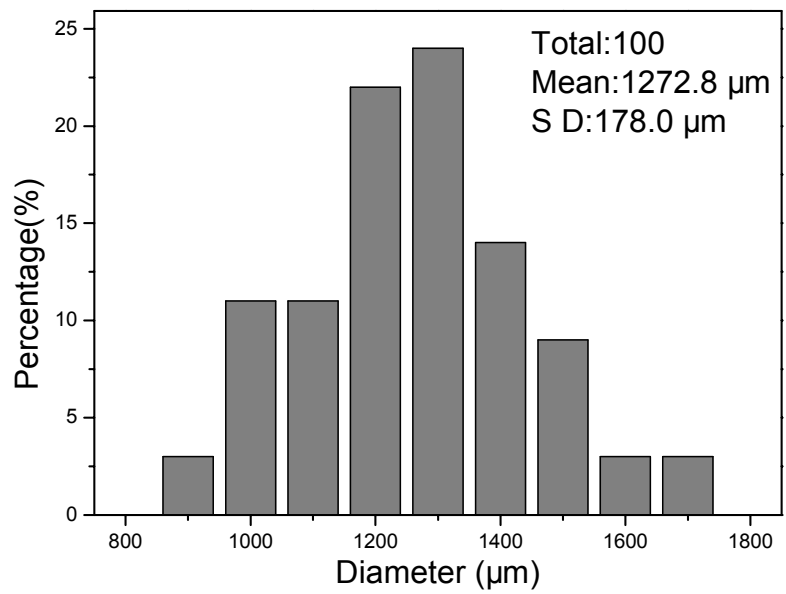

**Fig. S5.** Diameter distributions of the spheres before and after the CNT growth.

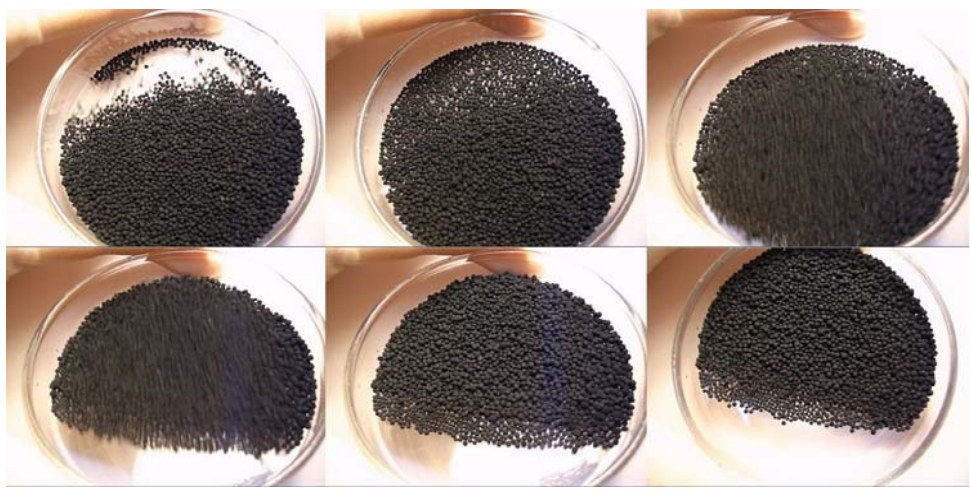

**Fig. S6.** Good fluidity of the as-grown spheres. They can easily move by tilting the gas dish. These pictures are captured from a video

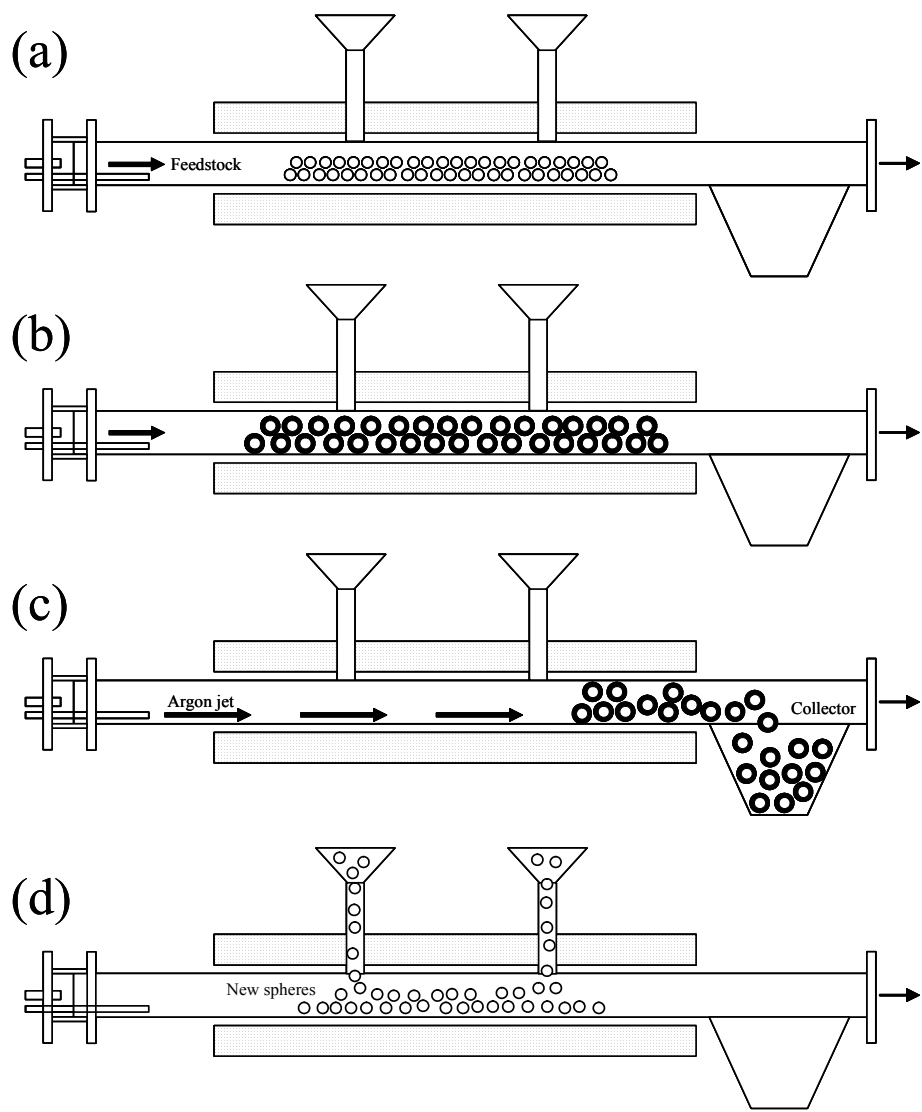

**Fig. S7.** Schematic diagram of a promising approach for continuous production of aligned CNT arrays on spheres: (a) a large number of spheres are first placed in the center of the reactor; (b) after feeding the carbon source and catalyst for a certain time, well aligned carbon nanotube carpets are formed vertically on all the spheres; (c) we can stop the feedstock and introduce a strong jet of argon to purge the as-grown spheres out of the hot zone to the collector; (e) after all the spheres are removed out of the reactor, new spheres can be supplied, then the reaction can be restarted and the whole process can be recycled.

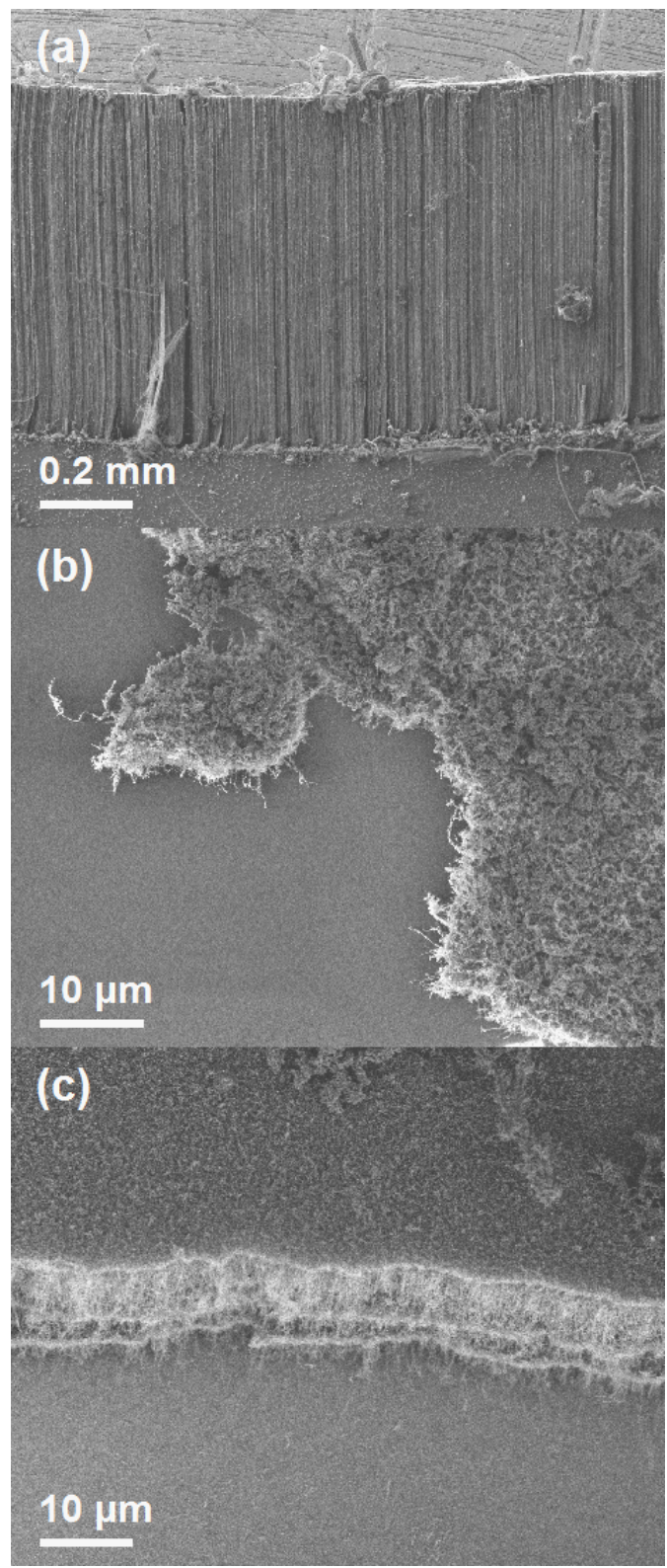

**Fig. S8.** SEM micrographs of as-grown product on pure oxide substrate: (a) SiO$_2$; (b) Al$_2$O$_3$; (c) ZrO$_2$